# On The Effective Rate and Error Rate Analysis over Fluctuating Nakagami-*m* Fading Channel

[1]Manpreet Kaur, [2]Puspraj Singh Chauhan, [3]Sandeep Kumar*, [4]Pappu Kumar Verma

[1,3]*Central Research Laboratory, BEL, Ghaziabad, India*
[2]*Dept of ECE, Pranveer Singh Institute of Technology, Kanpur, India,*
[4]*Dept of ECE, REC, Sonbhadra, India,*

*Corresponding author (Sandeep Kumar) mail id: *sann.kaushik@gmail.com*

**Abstract:** This paper provides a detailed analysis of the important performance metrics like effective capacity (EC) and symbol error rate (SER) over fluctuating Nakagami-*m* fading channel. This distribution is obtained from the ratio of two random variables, following the Nakagami-*m* distribution and the uniform distribution. Our study derives exact analytical expressions for the EC and SER under different modulation schemes, considering the effect of channel parameters. Recognising the importance of additive Laplacian noise (ALN) in today's scenario, it has been considered for the error performance analysis of the system. Numerical results are provided to validate the analytical findings, showing the influence of the channel conditions on the system performance. The study highlights that the EC and SER are significantly affected by the fading parameters, with higher fading severity and fluctuation rates lead to degraded performance. This work may be utilised for the design and optimization of the systems operating in environments characterized by fluctuating Nakagami-*m* fading.

**Keywords:** 6G, Composite fading, effective rate, Laplacian noise, *M-ary*-PSK, performance analysis, shadowing

## 1. Introduction

The behaviour of the wireless channel during the transmission of electromagnetic waves depends on various factors, which, in turn, impact the system's performance. Therefore, accurately characterizing wireless channels is essential for developing reliable wireless communication systems [1]. In the simplify ature, several distributions are used to model the channel behaviour, each characterizing different aspects of fading phenomena. Multipath fading is experienced when the signal travels along multiple paths to reach its destination, while shadowed fading occurs when major obstructions, such as hills and buildings, block the signal's primary path [2]. Multipath fading models capture various aspects of the channel, like non-linearity, multi-path components, LOS/NLOS etc., while shadowing models account for blockage due to obstacles. In the literature, several distributions were used to model multipath fading (Rayleigh, Ricean and generalised distributions etc.) and shadowing (Lognormal, inverse Nakagami-*m*, Gamma, and Inverse Gamma distributions etc.) [3-4].

A real-world communication scenario is modeled using composite channels that incorporate both multipath fading and shadowing concurrently. Consequently, the performance analysis of these composite channels is highly valuable both in theoretical research and practical applications [3]. To address the composite fading phenomenon, numerous statistical composite fading distributions have been proposed, leading to various performance analyses over these channels [5-8]. For instance, the expressions of physical layer performance metrics over κ-μ shadowed model are derived [5]. Similarly, leveraging the unique properties of composite fading channels, the authors developed a method for evaluating the maximum data rate over $F$ channel [6]. The authors in [7] created a framework to analyze key performance metrics over generalised inverse Gaussian distributions. Similarly the channel capacity over BX-shadowed fading model, under various policies, is provided in [8].

Advances in communication systems necessitate the introduction of new fading models, particularly for future-generation networks and emerging technologies [9]. To address the need for flexible and compatible fading models, the fluctuating Nakagami-$m$ distribution has been proposed to represent multipath components and blockage effects [10]. In this distribution, the Nakagami-$m$ component is influenced by the uniform distribution. Contrarily, the shadowing effects cause significant variations in signal strength and the use of the uniform distribution results in straightforward, closed-form PDF, which simplifies the analysis. Motivated by this, we have used uniform distribution for the characterization of shadowing effect in wireless communication channels.

To validate the usability of fluctuating Nakagami-$m$ distribution in emerging wireless systems, a comparison of measurement data in underwater acoustic communications (UAC), vehicle-to-vehicle (V2V) communications, and device-to-device (D2D) communication environment was carried out in [10]. The fluctuating Nakagami-$m$ model has shown to fit better than other models to show its flexibility and compatibility in describing various measurements scenarios [10].

In the Shannon's capacity theorem, delay is not considered as a QoS parameter. However, modern wireless communication systems often involve real-time applications such as IoT, D2D, BAN etc. For these applications, along with throughput, delay is also considered as a critical QoS requirement. To evaluate the system performance considering delay and reliability, the concept of effective capacity (EC) has been introduced in [11-12].

Effective rate performance over different fading channels has been conducted in literature under different communication scenarios [11-18]. In [12], the EC analysis over generalised fading channel was performed. The effective throughput over generalised gamma composite fading channels was explored in [13] and [14], respectively. The studies on EC performance over generalized composite fading channels and double shadowed κ-μ fading channels were presented in [15] and [16], respectively. The EC performance over α-η-κ-μ fading channels was examined in [17]. Analytical expressions for the EC of NOMA fading channels were derived in [18], while the EC with emphasis on reliability, taking into account power allocation effects, was examined in [19]

Long-distance transmission noise has an impulsive behaviour that can be characterised in the form of the additive Laplacian noise (ALN). In reality, simulating impulsive noise is very common in signal detection, signal processing, and communication studies [20]. The impact of ALN on wireless transmission over indoor, ultra-wide bandwidth (UWB), outdoor, underwater communications, visible light communication and multi-user interference systems has been presented in [20-26]. The performance of VLC in the presence of Gaussian and Laplacian noise has been considered in [20]. In another study, the effect of ALN in uplink NOMA systems has been explored in [21]. Bartoli *et. al.* investigated the data performance with reference to reliability in the harsh wireless environments affected by impulsive noise [22]. The study in [23] uses an ALN model to analyze the performance of the binary phase shift keying (PSK) in an underwater acoustic channel environment. In [24], the BER of binary PSK modulation over additive white generalized LN (AWGLN) channels is analyzed. In [25], the authors have derived the error probability for binary PSK, quadrature PSK, over extended generalized-K, whereas for *M-ary* PSK over generalised fading channels has been derived in [26].

Despite the many advantages of the fluctuating Nakagami-*m* model, still researchers have not explored it in diverse performance metrics. In addition, the effective throughput and symbol error rate (SER) performance with ALN over fluctuating Nakagami-*m* channel has not been investigated. The fluctuating Nakagami-*m* model holds significant potential in various emerging scenarios, particularly in UAC, V2V communications, and D2D communications. Motivated by this, we examine the EC performance over fluctuating Nakagami-*m* fading channels in this work. We further, examined the error performance with *M-ary* PSK modulation format with ALN. We also provide asymptotic analysis for the aforementioned performance indicators. Rather than having complex structure of SER under exact analysis, these expressions have simple functions. At last all the results have been dually corroborated with Monte-Carlo simulation to dig into the preciseness of the derived formulations.

The structure of paper is outlined as follows: In Section 2, we have considered the system model and describe the various components associated with it. Further, we defined the mathematical structure of the fluctuating Nakagami-*m* fading model. Section 3 describes the effective throughput analysis of the channel. Furthermore, in section 4 we discussed the evaluation of SER when communication system experiences ALN. Section 5 outlines the results on effective capacity and average SER. Finally, Section 6 incorporates concluding remarks.

## 2. System and Channel Model

In a communication system, the information is transmitted from source to destination. This information symbol is modulated before its transmission and is composed of two components, which describes their relationships with the amplitude and phase modulated carrier. The PDF of signal envelope following fluctuating Nakagami-*m* distribution is given by [10] as

$$f_R(r) = \frac{2m_s}{\Gamma(m)} r^{-1} G_{1,2}^{1,1}\left[\frac{m_s m}{(m_s-1)\Omega} r^2 \middle| \begin{array}{c} 1-m_s \\ m, -m_s \end{array}\right], \quad (1)$$

in which $m_s$ represents the severity of shadowing, $m$ represents the number of multipath clusters and $\Omega$ controls the mean total signal power of the PDF. The functions, $G_{p,q}^{m,n}(.)$ denotes the Meijer-G function and $\Gamma(.)$ represents the gamma function. At a given point, the received signal envelope, *R* of the propagation environment is given as

$$R = \sqrt{\sum_{i=1}^{m} U^{\frac{-1}{m_s}} X_i^2 + \sum_{i=1}^{m} U^{\frac{-1}{m_s}} Y_i^2}, \quad (2)$$

in which $X_i$ and $Y_i$ are independent Gaussian processes, with $E[X_i] = E[Y_i] = 0$ and $E[X_i^2] = E[Y_i^2] = \sigma^2$. The random variable *U* follows a uniform distribution. The relationship between $\Omega$ and $\sigma$ is defined as $\Omega = E[R^2] = 2\sigma^2 m\left(\frac{m_s}{m_s-1}\right)$.

The PDF of the $\gamma$ can be evaluated from the envelope distribution by following the conversion $\gamma = \frac{\bar{\gamma} R^2}{\Omega}$, with $\bar{\gamma}$ as the average received SNR and $f_\gamma(\gamma) = f_R\left(\sqrt{\frac{\Omega \gamma}{\bar{\gamma}}}\right) \Big/ 2\sqrt{\frac{\Omega}{\bar{\gamma}\gamma}}$ [27]. Thus, the expression of PDF of $\gamma$ following fluctuating Nakagami-*m* distribution is given by

$$f_\gamma(\gamma) = \frac{m_s}{\Gamma(m)} \gamma^{-1} G_{1,2}^{1,1}\left[\frac{m_s m}{(m_s-1)\bar{\gamma}} \gamma \middle| \begin{array}{c} 1-m_s \\ m, -m_s \end{array}\right]. \quad (3)$$

## 3. EC Analysis

The EC model determines the maximum data rate a channel can transmit while meeting specified delay constraints. Thus, EC is useful for analyzing system performance with delay bounds.

### 3.1 Exact Analysis

For a fading channel, the EC can be defined in [13] as

$$R(\bar{\gamma},\theta) = -\frac{1}{A}\log_2\left[E\left[(1+\gamma)^{-A}\right]\right] = -\frac{1}{A}\log_2\left[\int_o^\infty (1+\gamma)^{-A} f_\gamma(\gamma) d\gamma\right], \quad (4)$$

in which $A = \theta TB/\ln(2)$ represents the delay constraint. The symbols $B$, $T$ and $\theta$ represent the bandwidth, time duration of each frame and delay exponent respectively The system with higher $\theta$ signifies stringent QoS requirement, whereas a lower $\theta$ value signifies lenient QoS requirement and when $\underset{\theta \to 0}{EC} = Shannon's\ Ergodic\ capacity$ [28].

Substituting (3) in (4), and applying [29, eq. (10)], the expression of the EC over fluctuating Nakagami-$m$ fading channel is written as

$$R(\bar{\gamma},\theta) = -\frac{1}{A}\log_2\left[\frac{m_s}{\Gamma(m)\Gamma(A)}\int_o^\infty \gamma^{-1} G_{1,2}^{1,1}\left[\frac{m_s m}{(m_s-1)\bar{\gamma}}\gamma \bigg| \begin{matrix} 1-m_s \\ m,-m_s \end{matrix}\right] G_{1,1}^{1,1}\left[\gamma \bigg| \begin{matrix} 1-A \\ 0 \end{matrix}\right] d\gamma\right]. \quad (5)$$

Applying [30, eq. (2.24.1.1)], and using mathematical simplification, the final expression of EC is evaluated as

$$R(\bar{\gamma},\theta) = -\frac{1}{A}\log_2\left[\frac{m_s}{\Gamma(m)\Gamma(A)} G_{2,3}^{2,2}\left[\frac{m_s m}{(m_s-1)\bar{\gamma}}\gamma \bigg| \begin{matrix} 1,1-m_s \\ A,m,-m_s \end{matrix}\right]\right]. \quad (6)$$

### 3.2 Asymptotic Analysis

It helps us to understand the system's behaviour under both low-SNR and high-SNR regimes.
*a) High SNR*

The EC at high SNR ($\gamma \to \infty$) may be written using $(1+\gamma)^{-A} \cong \gamma^{-A}$ in equation (4) as

$$R^\infty(\bar{\gamma},\theta) \cong -\frac{1}{A}\log_2\left[\int_o^\infty \gamma^{-A} f_\gamma(\gamma) d\gamma\right]. \quad (7)$$

Substituting (3) in (7), we have

$$R^\infty(\bar{\gamma},\theta) \cong -\frac{1}{A}\log_2\left[\frac{m_s}{\Gamma(m)}\int_o^\infty \gamma^{-A-1} G_{1,2}^{1,1}\left[\frac{m_s m}{(m_s-1)\bar{\gamma}}\gamma \bigg| \begin{matrix} 1-m_s \\ m,-m_s \end{matrix}\right] d\gamma\right]. \quad (8)$$

Applying [29, eq. (24)], and using some mathematical simplification, the above expression can be simplified as

$$R^{\infty}(\bar{\gamma},\theta) \cong -\frac{1}{A}\log_2\left[\frac{m_s\Gamma(m-A)}{(m_s+A)\Gamma(m)}\left(\frac{m_s m}{(m_s-1)\bar{\gamma}}\right)^A\right]. \qquad (9)$$

*b) Low SNR*

In some scenarios, it is beneficial to derive the low-SNR regime of EC. Analyzing low SNR conditions based on average received SNR can lead to inaccurate results. Therefore, when the system operates under low SNR conditions, EC is evaluated in terms of $E_b/N_0$, as presented in [31].

$$R\left(\frac{E_b}{N_0}\right) \approx S_0 \log_2\left(\frac{E_b/N_0}{E_b/N_{0\,\min}}\right), \qquad (10)$$

in which

$$S_0 = \frac{-2(R'(0,\theta))^2 \ln(2)}{R''(0,\theta)}, \text{ and } \left(\frac{E_b}{N_0}\right)_{\min} = \frac{1}{R'(0,\theta)}, \qquad (11)$$

in which $R'(0,\theta)$ and $R''(0,\theta)$ can be mathematically represented as

$$R'(0,\theta) = \frac{E[\gamma]}{\ln(2)}, \qquad R''(0,\theta) = \frac{1}{\ln(2)}\left[A(E[\gamma])^2 - (A+1)E[\gamma^2]\right]. \qquad (12)$$

Substituting (3) in the definition of $E[\gamma]$, $E[\gamma^2]$ and considering [30, eq. (2.24.2.1)], we get

$$E[\gamma] = \frac{m_s\Gamma(m+1)}{(m_s-1)\Gamma(m)}\left(\frac{(m_s-1)\bar{\gamma}}{m_s m}\right). \qquad (13)$$

$$E[\gamma^2] = \frac{m_s\Gamma(m+2)}{(m_s-2)\Gamma(m)}\left(\frac{(m_s-1)\bar{\gamma}}{m_s m}\right)^2. \qquad (14)$$

Substituting (13) and (14) in (12), we get the closed-form expressions of $R'(0,\theta)$ and $R''(0,\theta)$, which can be used to get the results in (11). Further, the derived expressions of (11) are used to obtain the final expression of EC for low-SNR.

## 4. SER Analysis over Laplacian Noise

The SER is an important metric to analyze the quality of signal transmission through wireless channels. Here, the information symbol is modulated according to *M-ary* PSK (M-PSK) constellation, where M denotes the modulation order. Each symbol is comprised of two components, which describes their relationships with the amplitude and phase modulated carrier. In the MPSK scheme, symbols are evenly distributed across a circle with the assumption that all symbols have an equal probability of transmission.

The received signal is represented in terms of transmitted symbol, channel gain and noise. The noise considered in this case is ALN with PDF defined as [26, eq. (2)]

$$f_W(w) = \frac{1}{\sqrt{2\sigma_N^2}} \exp\left(-\sqrt{2}\frac{|w-m_N|}{\sigma_N}\right), \tag{15}$$

in which $m_N$ and $\sigma_N$ are the mean and variance of ALN.

## 4.1 Exact Analysis

Generally, the average SER, $\bar{P}(e)$ can be obtained by averaging the instantaneous SER, $P(e)$ and is given as

$$\bar{P}(e) = \int_0^\infty P(e) f_\gamma(\gamma) d\gamma. \tag{16}$$

*a) M-PSK Modulation*

The instantaneous SER of M-PSK modulation over ALN is expressed as

$$P(e) = \frac{8}{M} \sum_{l=0}^{\frac{M}{4}-1} g(l,\gamma) + \frac{2\tan\left(\frac{\pi}{M}\right)^2}{M\left(1-\tan\left(\frac{\pi}{M}\right)^2\right)} \exp(-2\sqrt{\gamma}), \tag{17}$$

where

$$g(l,\gamma) = \frac{1}{2\cos\left((2l+1)\frac{2\pi}{M}\right)} \left( \begin{array}{c} \cos\left((2l+1)\frac{\pi}{M}\right)^2 \exp(-k_2) \\ -\sin\left((2l+1)\frac{\pi}{M}\right)^2 \exp(-k_3) \end{array} \right)$$

$$- \frac{\sin\left(\frac{\pi}{M}\right)}{8\left(\cos\left(\frac{2\pi}{M}\right)+\sin\left(\frac{4l\pi}{M}\right)\right)} \exp(-k_4), \tag{18}$$

where $k_2 = \dfrac{2\sin\left(\dfrac{\pi}{M}\right)\sqrt{\gamma}}{\cos\left((2l+1)\dfrac{\pi}{M}\right)}$, $k_3 = \dfrac{2\sin\left(\dfrac{\pi}{M}\right)\sqrt{\gamma}}{\sin\left((2l+1)\dfrac{\pi}{M}\right)}$ and $k_4 = 2\sqrt{2}\cos\left(\dfrac{2l\pi}{M} - \dfrac{\pi}{4}\right)\sqrt{\gamma}$.

Substituting (17) in to (16), the average SER can be obtained as

$$\bar{P}(e) = \frac{8}{M}\sum_{l=0}^{\frac{M}{4}-1}\int_0^\infty g(l,\gamma)f_\gamma(\gamma)d\gamma + \frac{2\tan\left(\dfrac{\pi}{M}\right)^2}{M\left(1-\tan\left(\dfrac{\pi}{M}\right)^2\right)}\int_o^\infty \exp\left(-2\sqrt{\gamma}\right)f_\gamma(\gamma)d\gamma. \quad (19)$$

Both the integral terms in (19) can be written as the product of $f_\gamma(\gamma)$ and $\exp\left(-z\sqrt{\gamma}\right)$. So, to get the solution of (19), we need to solve (20)

$$I(z) = \frac{m_s}{\Gamma(m)}\int_o^\infty \gamma^{-1} G_{1,2}^{1,1}\left[\frac{m_s m}{(m_s-1)\bar{\gamma}}\gamma \,\bigg|\, \begin{matrix}1-m_s\\m,-m_s\end{matrix}\right]\exp\left(-z\sqrt{\gamma}\right)d\gamma. \quad (20)$$

Using [29, eq. (11)] and [30, eq. (2.24.1.1)], (20) can be simplified as

$$I(z) = \frac{m_s}{\Gamma(m)} G_{2,2}^{2,1}\left[z\sqrt{\frac{(m_s-1)\bar{\gamma}}{m_s m}} \,\bigg|\, \begin{matrix}1-m,1+m_s\\0,m_s\end{matrix}\right]. \quad (21)$$

Utilizing the result of (21), the final expression for the SER for M-PSK can be evaluated as

$$\bar{P}(e) = \frac{8}{M}\sum_{l=0}^{\frac{M}{4}-1} G(l) + \frac{2\tan\left(\dfrac{\pi}{M}\right)^2}{M\left(1-\tan\left(\dfrac{\pi}{M}\right)^2\right)} I(2), \quad (22)$$

where

$$G(l) = \frac{1}{2\cos\left((2l+1)\dfrac{2\pi}{M}\right)}\left(\begin{matrix}\cos\left((2l+1)\dfrac{\pi}{M}\right)^2 I(k_2)\\ -\sin\left((2l+1)\dfrac{\pi}{M}\right)^2 I(k_3)\end{matrix}\right)$$
$$-\frac{\sin\left(\dfrac{\pi}{M}\right)}{8\left(\cos\left(\dfrac{2\pi}{M}\right)+\sin\left(\dfrac{4l\pi}{M}\right)\right)} I(k_4). \quad (23)$$

*b) QPSK Modulation*

The instantaneous SER of QPSK modulation over ALN is expressed as

$$P(e) = \frac{3}{4}\exp(-2\sqrt{\gamma}) + \sqrt{\gamma}\exp(-2\sqrt{\gamma}). \tag{24}$$

Substituting (24) in to (16), utilizing the result (21) in the first terms and applying [29, eq. (11)] and [30, eq. (2.24.1.1)], in the second terms, (24) can be evaluated as

$$\bar{P}(e) = \frac{3}{4}I(2) + \frac{1}{\Gamma(m)}\sqrt{\frac{m_s(m_s-1)\bar{\gamma}}{m}} G_{2,2}^{2,1}\left[2\sqrt{\frac{(m_s-1)\bar{\gamma}}{m_s m}} \,\middle|\, \begin{matrix} 1-m, 1+m_s \\ 0, m_s \end{matrix}\right]. \tag{25}$$

*c) BPSK Modulation*

The instantaneous SER of BPSK modulation over ALN is expressed as

$$P(e) = \frac{1}{2}\exp(-2\sqrt{\gamma}) \tag{26}$$

Substituting (24) in to (16) and utilizing the result of (21), the average SER of BPSK modulation over ALN is simplified as

$$\bar{P}(e) = \frac{1}{2}I(2) \tag{27}$$

## 4.2 Asymptotic Analysis

The mathematical expressions derived in this section are generally in the form of Meijer-G functions, making direct interpretation of system behavior in terms of fading parameters somewhat challenging. Asymptotic frameworks help reduce this complexity when analyzing system performance at high power levels. In recent years, asymptotic analysis has gained significant attention for providing simpler and more direct results for performance parameters [32-33]. In this section, we derive the SER expressions in high-SNR regimes, utilizing asymptotic analysis to illustrate the limiting behavior. Asymptotic analysis' role is to look how the system behaves when $\gamma \to \infty$.

*a) M-PSK Modulation*

By using the methodology as defined in [33], we get the asymptotic expression of *I(z)* as

$$I_{Asymp}(z) = \frac{m_s^{\left(\frac{m}{2}+1\right)}}{(m+m_s)}\left(\frac{m}{z^2(m_s-1)\bar{\gamma}}\right)^{\frac{m}{2}}. \tag{28}$$

The final asymptotic expression for MPSK can be readily evaluated by putting (28) in to (22) and (23).

*b) QPSK Modulation*

By following the same approach as used above, we get the asymptotic expression for QPSK as

$$\bar{P}_{Asymp}(e) = \frac{3}{4} I_{Asymp}(2) + \frac{1}{2^m (m+m_s)} \sqrt{\frac{m_s(m_s-1)\bar{\gamma}}{m}} \left(\frac{m_s m}{(m_s-1)\bar{\gamma}}\right)^{\frac{m}{2}}. \quad (29)$$

*c) BPSK Modulation*

By following the same approach as used above, we get the asymptotic expression for BPSK as

$$\bar{P}_{Asymp}(e) = \frac{1}{2} I_{Asymp}(2). \quad (30)$$

## 5. Numerical Results

Here, we will give a variety of numerical and simulation findings for various scenarios using the previously derived analytical equations. The simulations and the evaluated results are compared where Monte-Carlo simulation results are generated with iteration $>10^6$. The numerical analysis and simulation findings are highly consistent, confirming the validity of our derivations. Numerous combinations of fading parameters are taken into consideration in simulation and numerical analysis to analyze their respective effects on the functionality of the systems. Figure 1 shows the EC versus $\bar{\gamma}$ for various values of fading parameters. The throughput is shown to increase with higher $m_s$ and $m$ values. The analytical curves for the channel closely match the asymptotic curves, becoming more accurate as the $\bar{\gamma}$ increases. Figure 2 illustrates effective throughput versus $A$ for various values of fading parameters. It is evident that as $m_s$ and $m$ value rises and $A$ value decreases, the system's channel capacity improves. For large $A$ values, the impact of channel parameters variation on the data capacity performance is more pronounced than for lower $A$ values. For instance, at $m_s = 4$ and $A = 1$, an increase in $m$ from 2 to 4, results in an 8% increase in EC whereas for $A = 10$ the same change in $m_X$ leads to a 32% increase in EC. Similarly, for a fixed $m = 4$, increasing $m_s$ from 4 to 6, results in a 1% EC increase for $A = 1$ and a 3% increase for $A = 10$.

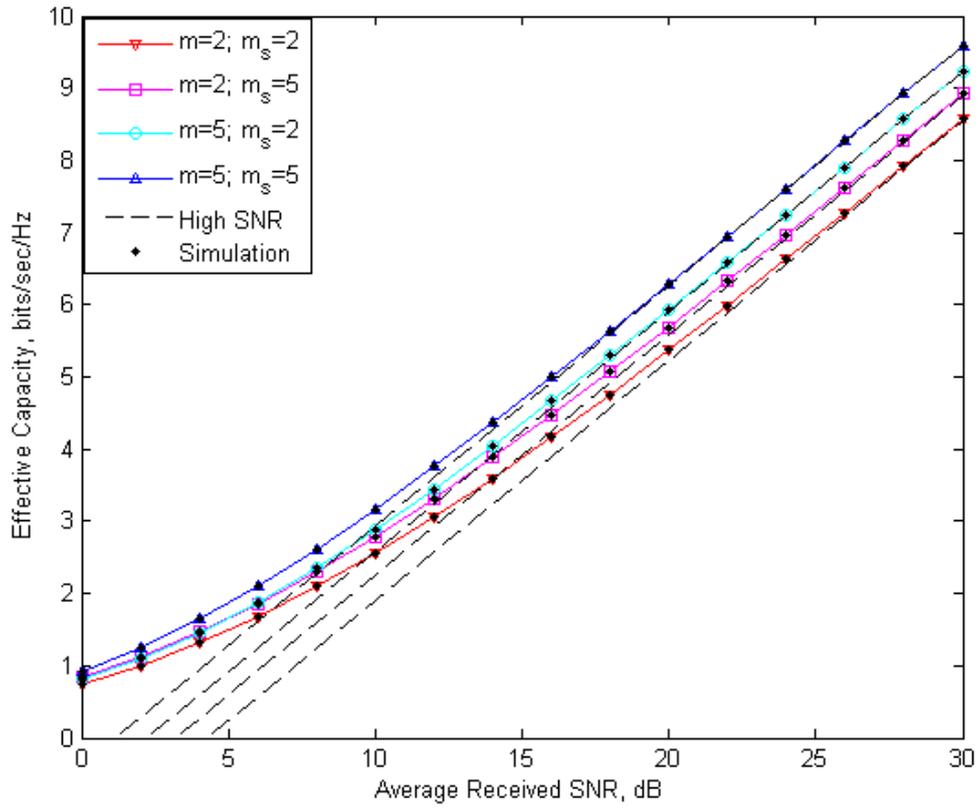

Figure 1 EC versus $\bar{\gamma}$ for different values of fading parameters.

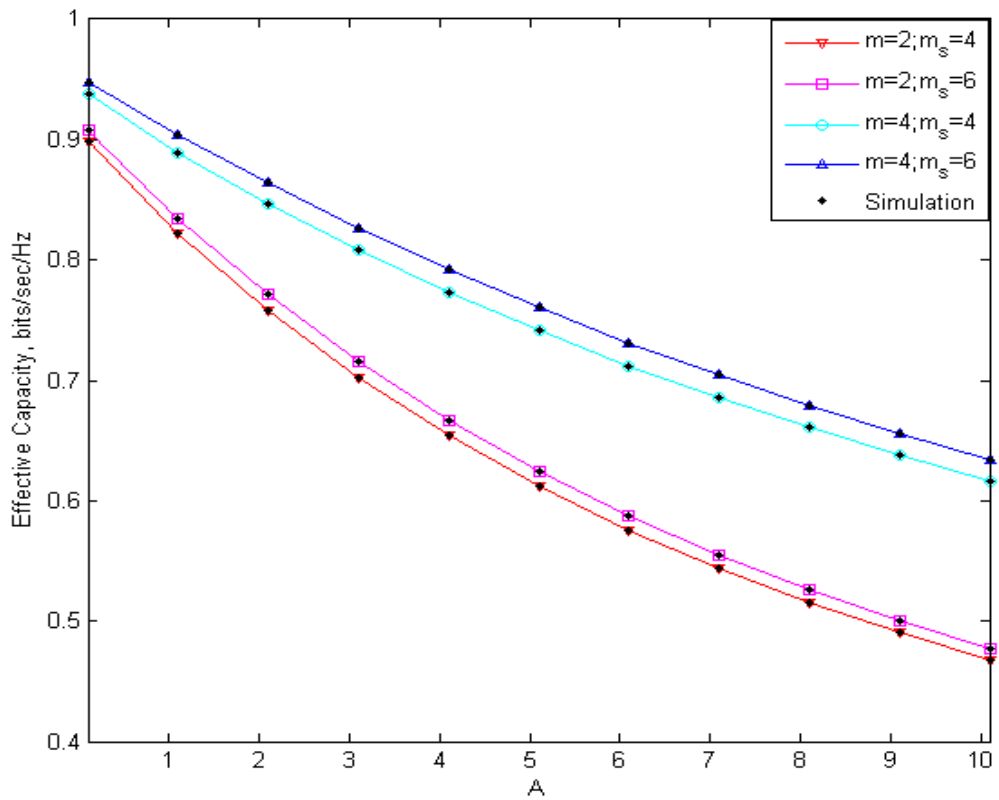

Figure 2 EC versus *A* for different values of channel parameters.

Figure 3 shows EC versus delay exponent for various $\bar{\gamma}$ values. System performance improves with higher $\bar{\gamma}$ values and lower delay exponent. The figure indicates that EC consistently decreases as the QoS exponent $\theta$ increases, implying that large delay constraint reduces the system's EC handling capability. For a fixed $m_s$ and $m$ with $\theta = 0.1$, EC increases by 40% when $\bar{\gamma}$ is increased from 5dB to 15dB, while for $\theta = 0.001$, EC increases by 65% for the same change in $\bar{\gamma}$. Figure 4 depicts the EC versus $E_b/N_0$ in the low SNR regime. The plot reveals that increasing the delay constraint $A$ degrades the EC performance. It is evident that EC decreases monotonically with increasing delay constraint $A$. Specifically, at $E_b/N_0 = 0$ dB, increasing $A$ from 1 to 5 results in a 42% reduction in EC. It is worth mentioning that any change in $A$, the minimum transmit SNR remains same, which is −6.7 dB in this scenario.

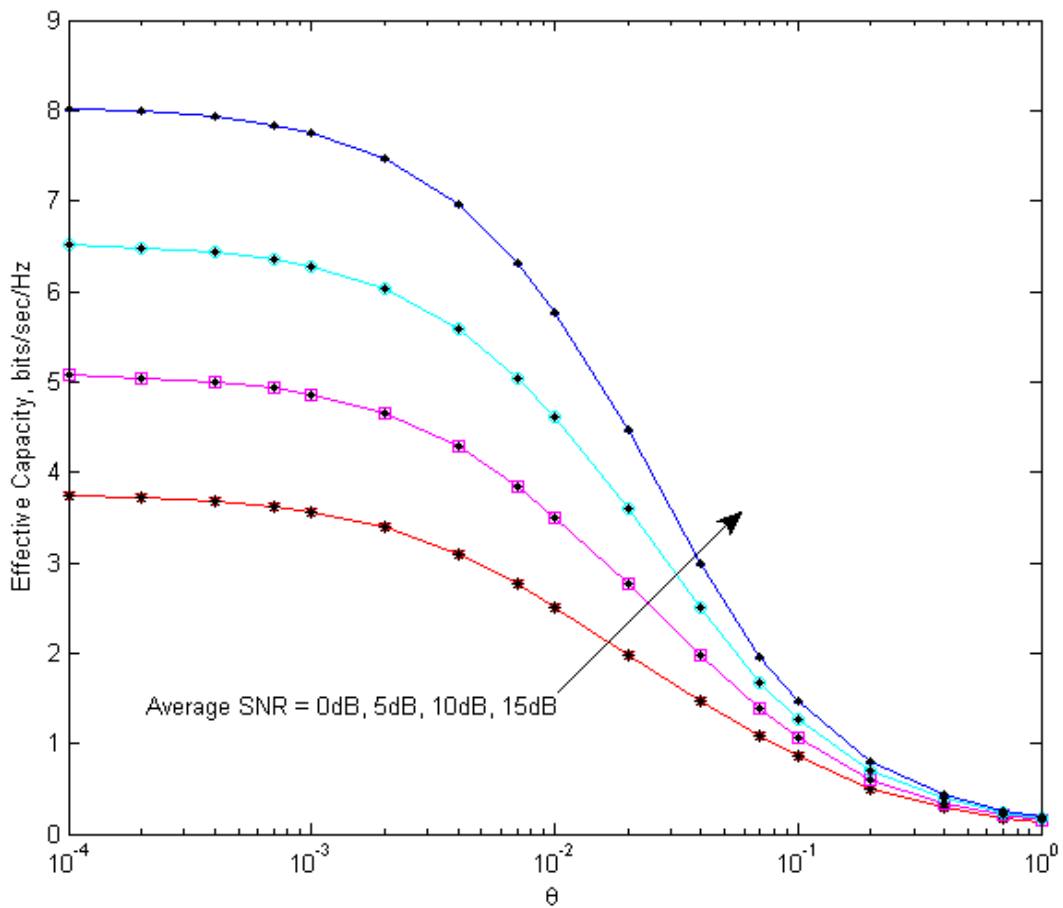

Figure 3 Graphical representation of EC versus delay exponent for different values of the $\bar{\gamma}$

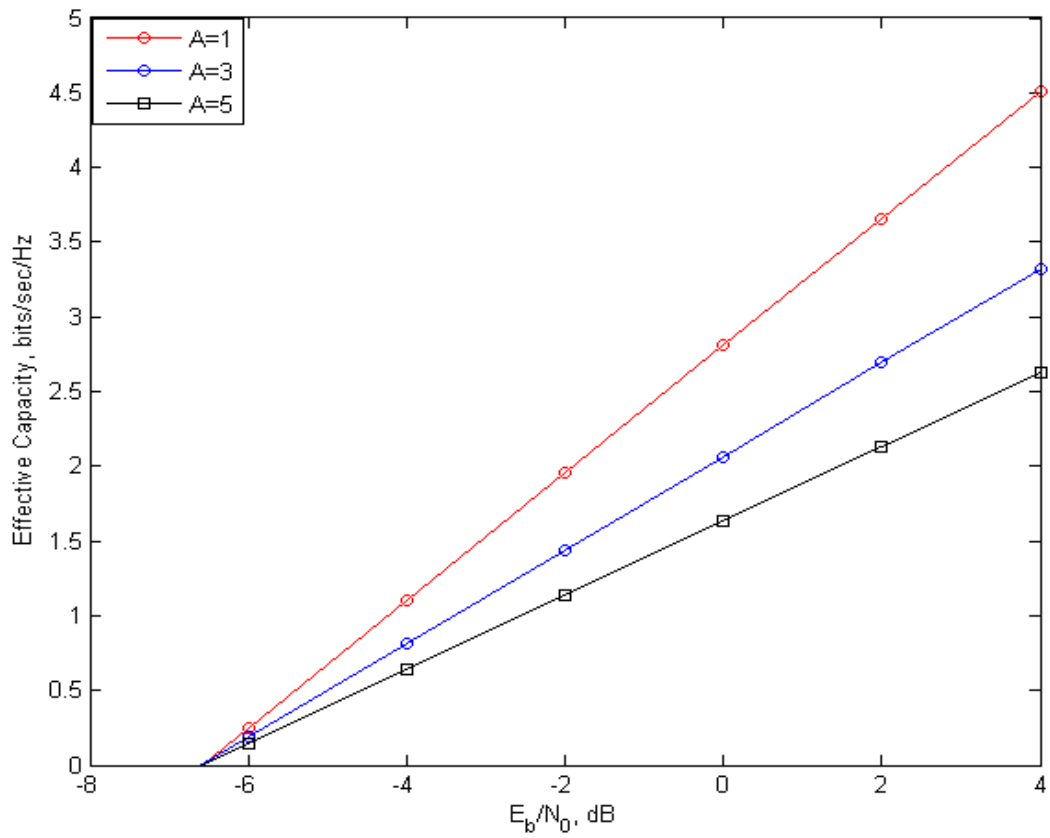

Figure 4 EC versus $E_b/N_0$ under low SNR regime for various values.

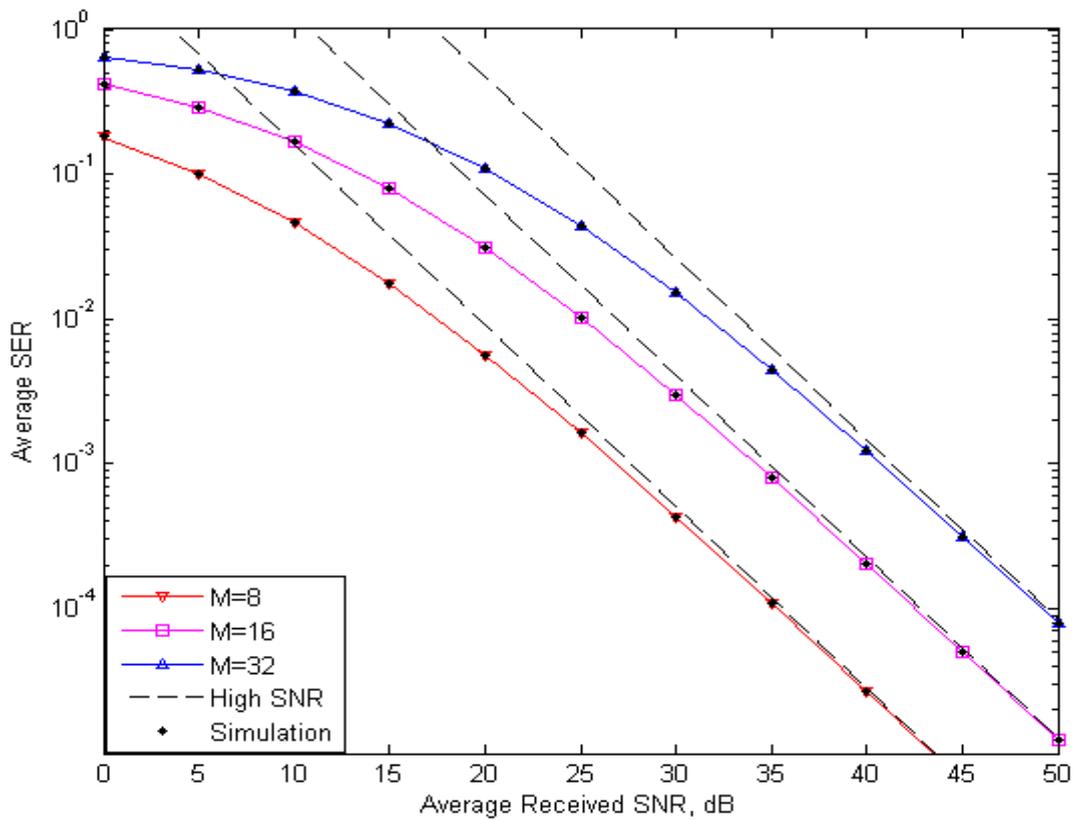

Figure 5 Average SER for *M*-PSK modulation with varying constellation size

In Figure 5, the average SER plot for *M*-PSK has been plotted against the average received SNR. The following are the system parameters: $m_s = m = 2.5$ and the constellation parameter M=8, 16, 32. The selection of these specific constellation sizes serves to validate the proposed formulations in equations (22), (23), and (28). The figure indicates that error probability rises with the increase in data transmission constellation size. This is because, as the constellation size grows, the distance between constellation points become closer, thereby increasing the likelihood of incorrect symbol prediction. From the plot, it is noted that reducing the constellation size from 32 to 16 results in an approximately 35 % improvement in uncertainty of error at $\bar{\gamma}$=0dB. By reducing it to 8, it exhibits an improvement of approximately 58 % compared to M=16 at $\bar{\gamma}$=0dB. It is additionally noted that the slope of the asymptotic plots is almost identical for all the constellation positions.

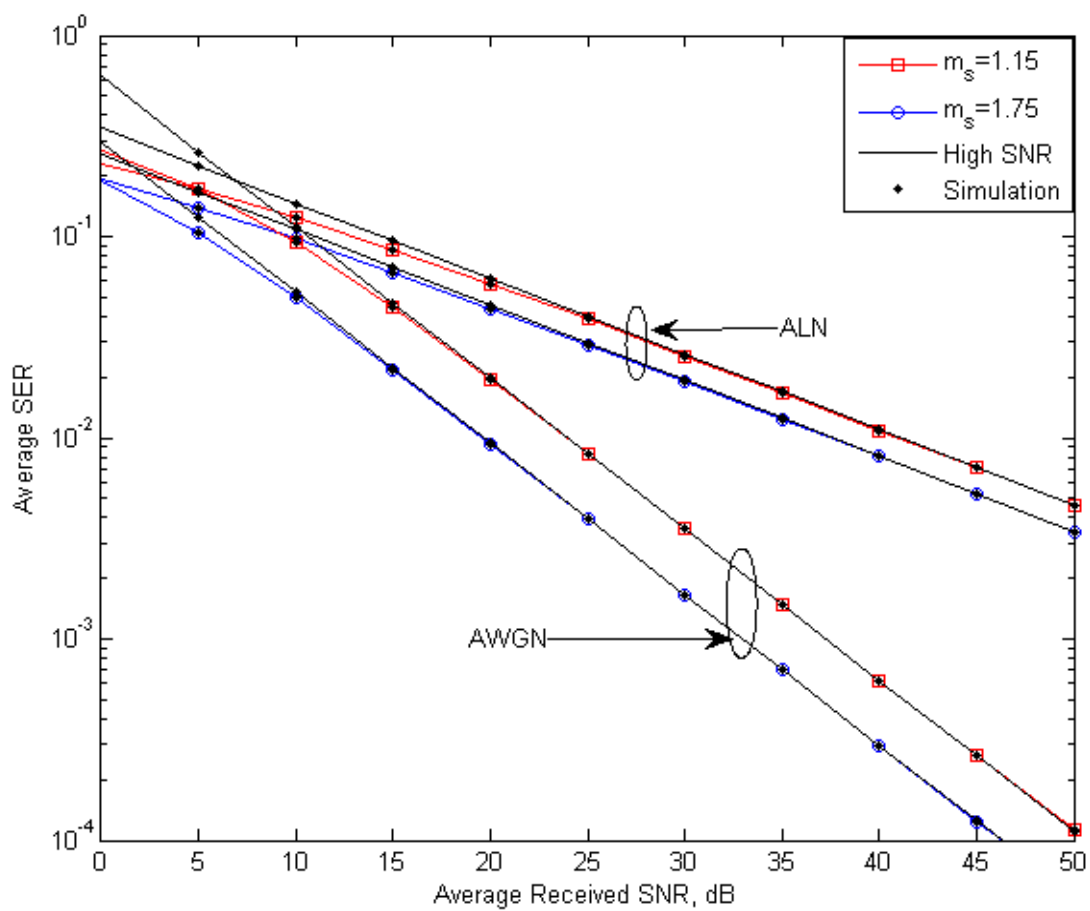

Figure 6 Average SER for binary PSK modulation with Gaussian and Laplacian with *m*=0.75

Figure 6 illustrates the average SER with binary PSK modulation against the average SNR. The figure compares error rates when the transmitted data is subject to Gaussian and Laplacian noise. The plot reveals that Laplacian noise significantly degrades system performance compared to Gaussian noise across a wide range of system parameters.

## 6. Conclusions

This paper has provided a comprehensive analysis of key performance metrics, namely effective capacity (EC) and SER over fluctuating Nakagami-*m* fading channels. By modeling the channel distribution as the ratio of two independent random variables we derived exact analytical expressions for EC and SER. Our analysis also incorporated the impact of ALN on the SER of the system under different modulation schemes, recognizing its relevance in contemporary communication scenarios. The numerical results validated our analytical findings, illustrating the significant influence of channel conditions on system performance. Overall, this work contributes valuable knowledge for the design and optimization of wireless systems in environments characterized by fluctuating Nakagami-*m* fading.

**Declarations:**

**Code availability** N/A

**Conflicts of Interest** There is no conflict of interest.

**Authors' contributions** All the authors have equally contributed in this manuscript.

**Data Availability** N/A (There is no research data outside the submitted manuscript file.)

**Funding** No funding was received.